\documentclass[fleqn,10pt]{wlscirep}
\usepackage{amsmath}
\title{Anderson Localization in Degenerate Spin-Orbit Coupled Fermi Gas with Disorder}

\author[1,2]{Sheng Liu}
\author[1,2] {Xiang-Fa Zhou}
\author[1,2]{Guang-Can Guo}
\author[1,2,*]{Yong-Sheng Zhang}

\affil{Key Laboratory of Quantum Information, University of Science and Technology of China, Hefei, 230026, China}
\affil[2]{Synergetic Innovation Center of Quantum Information and Quantum Physics, University of Science and Technology of China, Hefei, 230026, China}

\affil[*]{Corresponding author: yshzhang@ustc.edu.cn}

\begin{abstract}
Competition between superconductivity and disorder plays an essential role in understanding the metal-insulator transition. Based on the Bogoliubov-de Gennes framework, we studied an 2D $s$-wave fermionic optical lattice system with both spin-orbit coupling and disorder are presented. We find that, with the increase of the strength of disorder, the mean superconducting order parameter will vanish while the energy gap will persist, which indicates that the system undergoes a transition from a superconducting state to a gapped insulating state. This can be confirmed by calculating the inverse participation ratio. We also find that, if the strength of disorder is small, the superconducting order parameter and the energy gap will decrease if we increase the strength of spin-orbit coupling and Zeeman field. In the large disorder limits, increase the strength of spin-orbit coupling will increase the mean superconducting order parameter. This phenomenon shows that the system is more insensitive to disorder if the spin-orbit coupling is presented. Numerical computing also shows that the whole system breaks up into several superconducting islands instead of being superconductive.
\end{abstract}

\begin{document}
\flushbottom
\maketitle

\section*{Introduction}
Metal-insulator transition is a long standing topic in condensed matter physics. Understanding the mechanism of metal-insulator transition will be of great help in designing electronic devices. In 1949, Mott\cite{Mott:1949} proposed a simple model shows that metal-insulator transition can be induced by electron-electron interaction, and this kind of insulator that called  Mott insulator nowdays covers a lot of materials. The Mott's theory is a marvellous theory,  but it is not the whole story. In 1958, Anderson\cite{Anderson:1958} took a different angle to study this problem, he considered disorder in the system. He showed that, when the disorder is increased, the system will undergo a phase transition from metal to insulator. Later, in the famous paper written by the `gang of four' \cite{Abrahams:1979}, it was shown by using scaling theory that in 1-dimensional and 2-dimensional system there will always exist localization as long as the disorder presents, no matter how small the disorder is. It is a different case in 3-dimensional system that there exists a critical disorder strength, under which the system is in extended state while above it the system is in localized state.

Since the discovery in 1958, Anderson localization has been studied extensively in a lot of obviously different systems including ultracold quantum gases\cite{Shapiro:2012,Semeghini:2015,Sanchez:2007} that attracting a lot of attentions recently. Among them, it is amazing when the disorder is introduced into the superconductor, as superconductivity and Anderson localization have the opposite effects. In superconductors, electrons are bounded into Cooper pairs that then condense into a zero-resistance, macroscopic quantum state. In contrast, disorder induces localization of the electrons' wave function that will transform the metal into an insulator with diverging resistance.

Over the years, precise theoretical study of the Anderson model has not been conducted because the interaction in real system is not negligible. However, a lot of experiments\cite{Billy:2008,Modugno:2010} have been realized in ultracold quantum gases that the interaction has been tuned to be zero using Feshbach resonance and the localized wave function has been visualized. If attractive interaction between fermions are turned on in a system without disorder, this system will be in the superconducting state under a critical temperature $T_c$. This superconducting state can also exist in a 2-dimensional disordered system below a critical disorder even when all single-particle states are localized\cite{Sacepe:2011}. The superconductor-insulator transition has been studied very extensively\cite{Cai:2013,Cui:2008,Dubi:2007,Dubi:2008,Ghosal:1998,Ghosal:2001,He:2013}. One of the most important aspects of ultracold quantum gases is that the interaction can be tuned, which enable the realization of mobility edge in 3-dimensional quantum gases\cite{Semeghini:2015}.

The superconducting-localization transition in a system without spin-orbit coupling (SOC) has been studied\cite{Ghosal:1998,Ghosal:2001,Dubi:2007,Sacepe:2011,Roati:2008,Potirniche:2014} by other researchers. It is shown that, with the increase of disorder, the mean order parameter $\Delta=\frac{1}{N}\sum|\Delta_i|$ will decrease to zero. Surprisingly, the energy gap $E_g$ continues to show a finite value. This non-zero energy gap at larger disorder is due to the breakup of the system into several {\it superconducting islands}\cite{Ghosal:1998,Ghosal:2001}. The same system has also been investigated with spin-dependent disorder\cite{Nanguneri:2012,Jiang:2013}. It is found that, with the increase of the strength of disorder, the energy gap and the mean order parameter both approach to zero, which is significanly different from the system with spin-independent disorder. The scaling theory predicts that there will be an Anderson transition in $d>2$ dimensional systems which is lacked in 2D system. The system with SOC has been studied by several groups\cite{Sanchez:2008,Zhou:2013}. In Ref. [22], they found that SOC can lead to a mixing of localized states and extended states and lead to the appearance of mobility edge which indicates there will be an Anderson transition in 2D system.

\begin{figure}[ht]
\centering
\includegraphics[height=7cm]{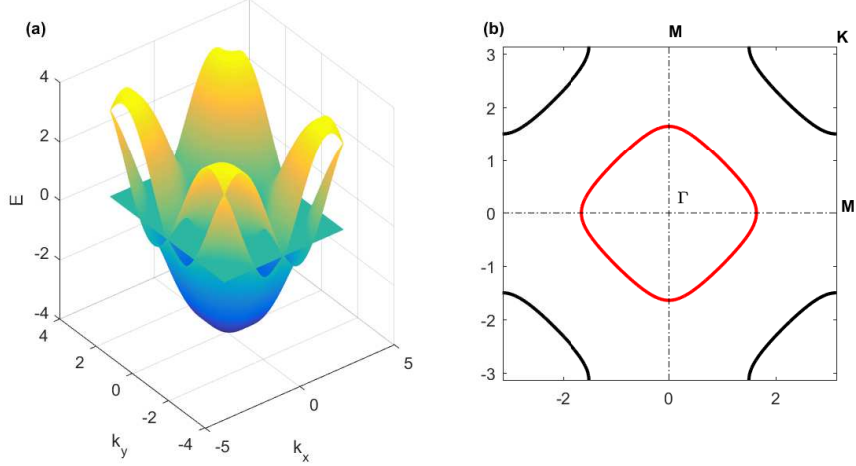}
\caption{(a) Energy bands of non-interacting system. (b) Zero energy Fermi surface (corresponding to half-filling). We set $h=0$ and $\mu=0$, and choose $\alpha=0.75$. In (a) we plotted the zero energy surface. In (b), the red circle is the intersection of zero energy plane and the lower Rashba band, the four black half-circles are the intersection of zero energy plane and the upper Rashba band.}
\label{fig:band}
\end{figure}
In this paper, we investigate the effect of SOC and Zeeman field along with spin-independent disorder on an $s-$wave superconductor in a 2D optical lattice defined by Eq. (1) in the section of Results, and analyze it in detail based on the Bogoliubov-de Gennes framework\cite{Gennes:1966,Ghosal:1998,Xu:2014}. Our goal is to find how the mean order parameter $\Delta$ and the energy gap $E_g$ vary in the presence of both SOC and disorder. We find that, with the increase of the strength of disorder, $\Delta$ tends to zero while $E_g$ remains a finite value which confirms the results in Ref. [13]. Meanwhile, the distribution of $\Delta$ broadens very much. When we increase the strength of SOC, it shows that the speed of the decreasing of $\Delta$ is reduced. We also find that $\Delta$ and $E_g$ are reduced if we fix the disorder strength while increasing the strength of SOC or Zeeman field which is consistent with Ref. [27]. In the absence of disorder and with small Zeeman field, increasing the strength of SOC reduces both $\Delta$ and $E_g$\cite{Sun:2013}. While for larger Zeeman field, increasing the strength of SOC only has large effect on $\Delta$ and small effect on $E_g$. To study the localization effect in this system, we also calculated the inverse participation ratio. It shows that, for small strength of SOC, the system will go to localized state with increasing disorder. And for large strength of SOC, with the increase of the strength of disorder, the system will go to a mixed state that consists of both extended states and localized states.

\section*{Results}
We consider fermionic cold atoms confined in a 2D optical lattice with Rashba-type SOC, an out-of-plane Zeeman field\cite{Xu:2014} and uniformly distributed disorder. The system's Hamiltonian is
\begin{equation}
H=H_0+H_{so}+H_z+V,
\end{equation}
which contains the single particle term
\begin{equation}
H_0=-t\sum_{\langle ij\rangle,\sigma}c_{i\sigma}^{\dagger}c_{j\sigma}-\mu \sum_{i\sigma}{\hat n}_{i\sigma}-U\sum_{i}{\hat n}_{i\uparrow}{\hat n}_{i\downarrow},
\end{equation}
the Rashba-type spin-orbit coupling term
\begin{equation}
\begin{aligned}
H_{so} ={} & \sum_{i}(-\alpha c_{i_{x+1}\uparrow}^{\dagger}c_{i\downarrow}+\alpha c_{i_{x+1}\downarrow}^{\dagger}c_{i\uparrow} +\alpha c_{i_{x-1}\uparrow}^{\dagger}c_{i\downarrow}-\alpha c_{i_{x-1}\downarrow}^{\dagger}c_{i\uparrow}\\
&+i\alpha c_{i_{y+1}\uparrow}^{\dagger}c_{i\downarrow}+i\alpha c_{i_{y+1}\downarrow}^{\dagger}c_{i\uparrow} -i\alpha c_{i_{y-1}\uparrow}^{\dagger}c_{i\downarrow}-i\alpha c_{i_{y-1}\downarrow}^{\dagger} c_{i\uparrow}),
\end{aligned}
\end{equation}
the out-of-plane Zeeman field
\begin{equation}
H_z=h\sum_{i}(c_{i\uparrow}^{\dagger}c_{i\uparrow}-c_{i\downarrow}^{\dagger}c_{i\downarrow}),
\end{equation}
and the disorder
\begin{equation}
V=\sum_{i}V_i {\hat n}_i.
\end{equation}

Here $t$  is the hopping energy, $\mu$ is the chemical potential, $U>0$ is the attractive interaction strength, $\alpha$ is the strength of SOC, $h$ is the strength of Zeeman field, $\langle ij\rangle$ denotes the summation is for the nearest-neighbour sites, $V_i$ is the disorder strength which is uniformly distributed in $[-W, W]$, and $i_{x\pm1}, i_{y\pm1}$ mean the hopping is occurred in the $x$ and $y$ direction, respectively. $\{c_{i\sigma},  c_{i\sigma}^{\dagger}, \sigma=(\uparrow, \downarrow)\}$ are the Fermion annihilation and creation operators. $\hat n_{i\sigma}=c_{i\sigma}^{\dagger}c_{i\sigma}$ is the number operator.

We employed the mean-field theory to decouple the interaction term which itself is quadratic. And then we introduced the quasi-particle operators to study the superconducting phase. The detailed derivation is in the Methods section. In our numerical calculation, we keep density fixed $\langle n \rangle = 0.875$ which is near half-filling, and hopping energy $t = 1.0$. We choose the on-site interaction strength to be $U=4.0$ which is large enough to ensure that the coherence length is within the system so our numerical calculation is trustful. In order to reduce the fluctuations caused by one single random on-site disorder, we obtain the results averaged over 15 disorder realizations. Finally, we set our system size to be $N=12\times 12$ and choose the periodic boundary condition to reduce the finite size effect.

\begin{figure}[ht]
\centering
\includegraphics[height=9cm]{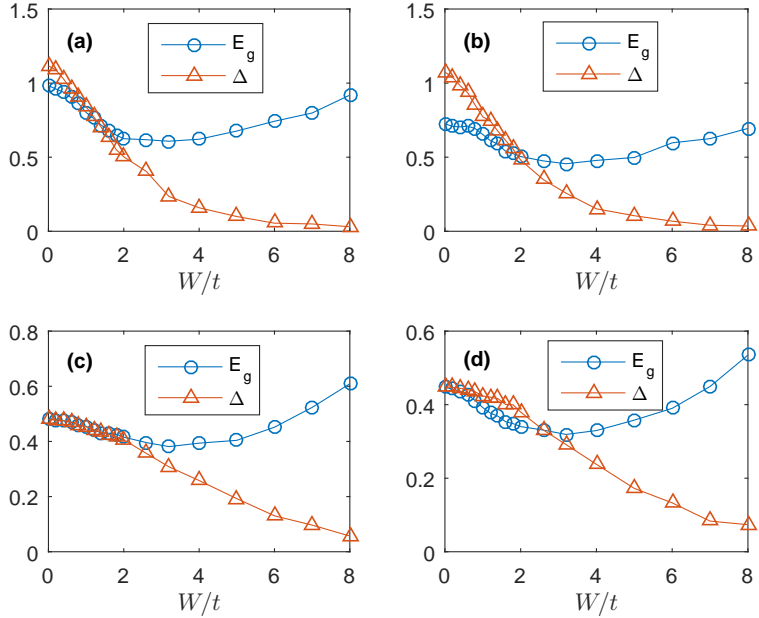}
\caption{Energy gap $E_g$ and mean order parameter $\Delta$ as a function of disorder width $W$. We choose four sets of parameters. (a) $\alpha=0.75,h=0.3.$ (b) $\alpha=0.75,h=0.6.$ (c) $\alpha=1.5,h=0.3.$ (d) $\alpha=1.5,h=0.6.$}
\label{fig:alpha}
\end{figure}

In Fig. \ref{fig:band} we calculated the energy bands for the non-interacting system (we set $h=0$ and $\mu=0$), the degenerated bands are lifted by SOC and the bands are
\begin{equation}
\begin{aligned}
E_{+}=-2t\cos{\alpha}(\cos{k_x}+\cos{k_y})+2t\sin{\alpha}\sqrt{\sin^2{k_x}+\sin^2{k_y}},\\
E_{-}=-2t\cos{\alpha}(\cos{k_x}+\cos{k_y})-2t\sin{\alpha}\sqrt{\sin^2{k_x}+\sin^2{k_y}}.
\end{aligned}
\end{equation}

In Fig. \ref{fig:band}(b), we plotted the zero energy Fermi surface for $\alpha=0.75$ at half filling. We see there is a particle Fermi pocket around $\Gamma$ and a hole Fermi pocket around $K$\cite{Sun:2013}. There are always two zero energy Fermi points located at $M$, this can be found if we let $E_{+}=E_{-}=0$. And we see that, as increasing of SOC, both the two pockets shrinks to Fermi points ($\Gamma$ and $K$) where the energy is zero.

Fig. \ref{fig:alpha} shows the evolution of the energy gap $E_g$ and mean order parameter $\Delta$ as functions of disorder width $W$ for different strength of SOC and Zeeman field. We find that, with the increase of the strength of disorder, $\Delta$ will vanish while $E_g$ remains a finite value. This is just like the system without SOC \cite{Ghosal:1998,Ghosal:2001}. In the absence of disorder, $E_g$ and $\Delta$ will decrease if we increase the strength of SOC or Zeeman field, respectively. This is in contrast different from other results\cite{Chen:2012}. In Ref. [28], they found that, as we increase SOC, the superconducting order parameter will increase. To understand this difference, we need the help of the energy bands of the non-interacting system. For the filling number $n=0.875$ we considered, as we increase SOC, the Fermi pockets around $\Gamma$ and $K$ both shrinks to Fermi points which are located exactly at points $\Gamma$ and $K$. This suppresses the density of states and reduces the pairing\cite{Sun:2013}. For small Zeeman field, the reducing is very significant for both $E_g$ and $\Delta$. For large Zeeman field, the reduction of $\Delta$ is significant while $E_g$ is barely affected by SOC.

Although the increase of the strength of SOC will reduce $\Delta$ and $E_g$, it is still different from the system without SOC. From Fig. \ref{fig:alpha}(a) and (c),  with the same Zeeman field $h=0.3$ but different SOC strength, we find that the mean order parameter $\Delta$ is more robust against the disorder for large SOC. With the same strength of disorder, $\Delta$ of system with large SOC is lower compared to the system with small SOC. This means that the superconducting density is lower than the system without SOC, but the superconducting capability is much enhanced by SOC.

\begin{figure}[ht]
\centering
\includegraphics[height=9cm]{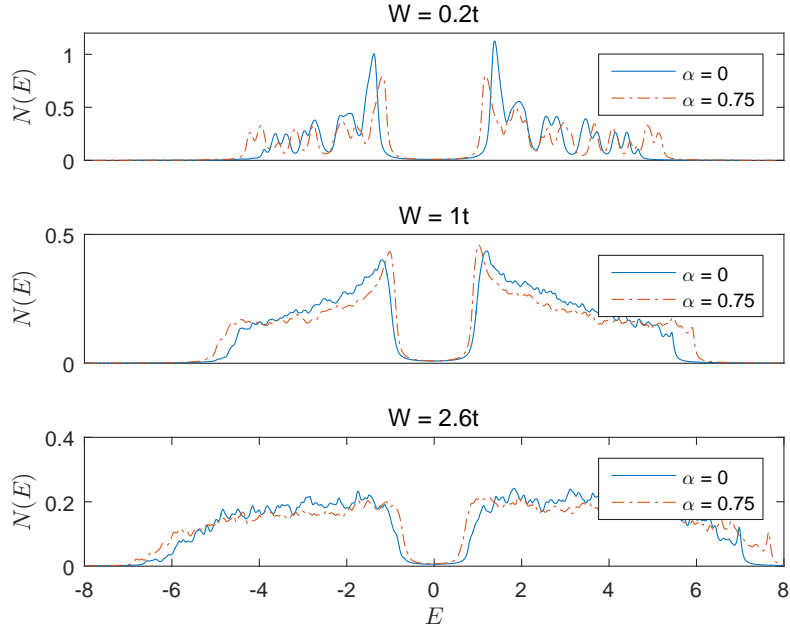}
\caption{Density of states. Zeeman field $h=0.3$. The corresponding strength of SOC are labelled in the figures.}
\label{fig:disorder_delta}
\end{figure}

In all the cases, the energy gap $E_g$ persists even at very high disorder strength in spite of growing number of sites with $\Delta_i=0$. They barely change in the whole range of disorder strength we have considered. Actually, we have $E_g=\frac{U}{2}$ if $W>>U,t$. Because, as the single-particle states become more localized as indicated by $\Delta$, the effective attraction is more enhanced, leading to a larger energy gap $E_g$\cite{Trivedi:2012}. To understand how the energy gap evolves as the local order parameters become highly inhomogeneous, we computed the disorder averaged density of states (DOS) as follows
\begin{equation}N(E) = \sum_{n,i,\sigma} [|u_{i\sigma}^{n}|^2\delta(E-E_n)+|v_{i\sigma}^{n}|^2\delta(E+E_n)].
\end{equation}
$\{E_n\}$ is the eigenenergy of the BdG matrix. In Fig. \ref{fig:disorder_delta} we ploted the DOS for different strength of disorder and SOC. From all the figures presented in Fig. \ref{fig:disorder_delta}, it can be seen that the energy gap will not vanish no matter SOC is presented or not. It indicates that the system undergoes a transition from a gapped superconductor to a gapped insulator\cite{Bouadim:2011}. Under small strength of disorder, there is a sharp peak near the energy gap, which indicates that the states are piled up near the energy gap and the system is in superconducting state. The reason is that, in the absence of disorder, energy gap $E_g$ equals the average of order parameter $\Delta$. When we increase the strength of disorder, the sharp peak is smeared out, but the energy gap remains finite. This means the system undergoes a transition to a gapped insulator which is also conformed by other group\cite{Cao:2015}. For large strength of disorder, the states of the system are distributed in the whole eigenenergy interval. Actually, SOC will broaden the energy distribution as we can see from Fig. \ref{fig:disorder_delta}. From this we can say that the whole system is not superconducting but forms several {\it superconducting islands}. This can be explained as follows: SOC and Zeeman field induce a mismatched Fermi surface, so spin-up (spin-down) particles with a particular momentum can not find spin-down (spin-up) partners to form a Cooper pair.  Hence there exist many paired and unpaired particles. Under the influence of disorder, the paired particles are localized and form superconducting islands.

Although SOC does not change the physical pictures qualitatively, it still causes some quantitative difference. As we can see from Fig. \ref{fig:disorder_delta}, SOC reduces the energy gap $E_g$ under the same conditions compared to the system without SOC.

\begin{figure}[ht]
\centering
\includegraphics[height=9cm]{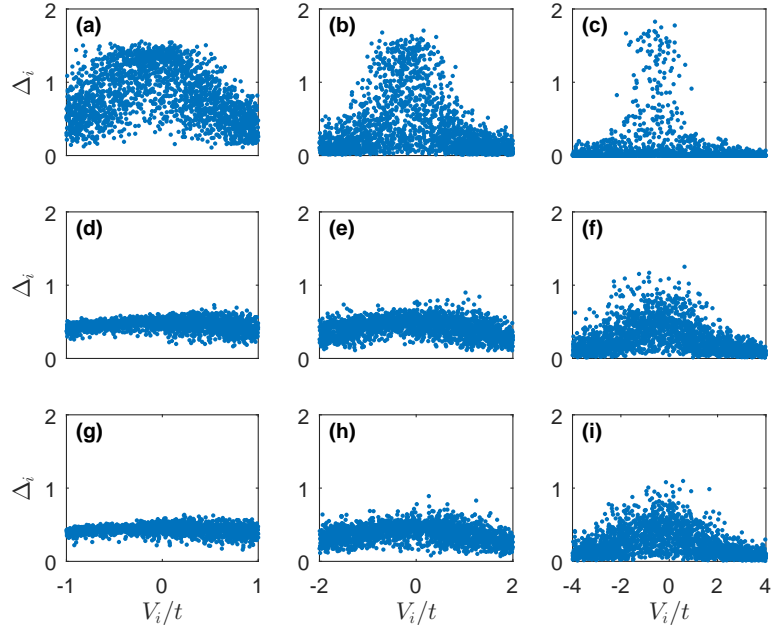}
\caption{Histogram for local superconducting order parameter $\Delta_i$ as a function of local lattice disorder strength $V_i$. (a) $\alpha=0,h=0,W=1.$ (b) $\alpha=0,h=0,W=2.$ (c) $\alpha=0,h=0,W=4.$ (d) $\alpha=1.5,h=0,W=1.$ (e) $\alpha=1.5,h=0,W=2.$ (f) $\alpha=1.5,h=0,W=4.$ (g) $\alpha=1.5,h=0.6,W=1.$ (h) $\alpha=1.5,h=0.6,W=2.$ (i) $\alpha=1.5,h=0.6,W=4.$}
\label{fig:disorder}
\end{figure}

\begin{figure}[ht]
\centering
\includegraphics[height=4.5cm]{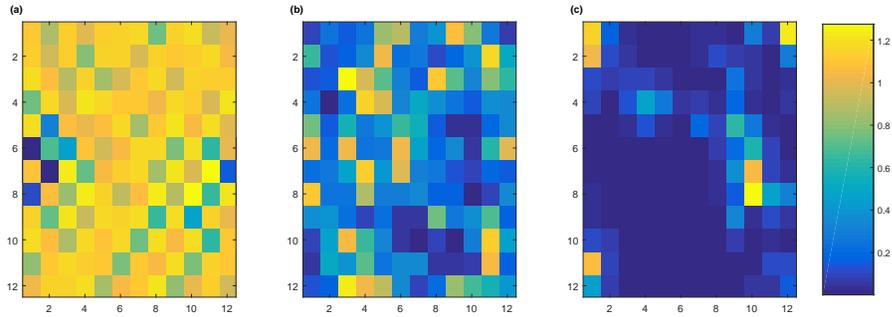}
\caption{Spatial distribution of local superconducting order parameters $\{\Delta_i\}$ for $\alpha = 0.75$ and $h=0.3$. (a) $W=0.2,$ (b) $W=2.0,$ (c) $W=4.0.$}
\label{fig:local_delta}
\end{figure}

As we can find from Fig. \ref{fig:alpha}, in the presence of SOC, $\Delta$ is more insensitive to the disorder.  In Fig. \ref{fig:disorder}, we plotted the histogram of local order parameter $\Delta_i$ as a function of local site disorder $V_i$. We first analyse the system without SOC and Zeeman field. When the strength of disorder is small ($W=1$),  the whole order parameters $\{\Delta_i\}$ are lowed compared to the disorder-free system. The maximum of $\Delta_i$ is smaller than the mean order parameter of the disorder-free system. On the sites with large disorder, the local order parameters are suppressed. The order parameters increase for sites with small absolute value $|V_i|$. When the strength of disorder is moderate ($W=2$), the order parameters with large disorder are suppressed significantly, some of them are zeros, which indicates that the superconducting is totally suppressed. In the condition of strong disorder ($W=4$), the fluctuations of local order parameter $\Delta_i$ in the region of small $|V_i|$ are very strong. There even exist some sites whose local order parameter are larger than the disorder-free system. The system breaks up into clusters with non-zero order parameters surrounded by some zero order parameter sites. This indicates the existence of {\it superconducting islands} as we plotted in Fig. \ref{fig:local_delta}. Although the disorder will always have the effect of breaking the superconducting pairing, but in the strong disorder limit, when two localized particles with opposite spin form a Cooper pair, the strong disorder will protect the Cooper pair from being affected by other particles outside the superconducting island. These kinds of Cooper pairs are only accumulated on the sites with strong disorder, this explains why the whole system is not superconducting while there exist some superconducting islands.

\begin{figure}[ht]
\centering
\includegraphics[height=9cm]{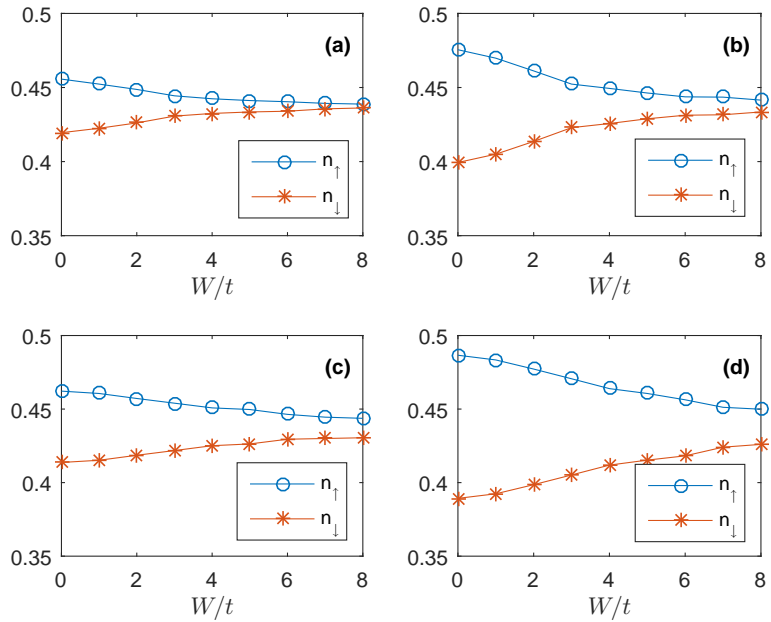}
\caption{Number density for spin-up and spin-down particles with different strength of SOC and Zeeman field. (a) $\alpha=0.75,h=0.3.$ (b) $\alpha=0.75,h=0.6.$ (c) $\alpha=1.5,h=0.3.$ (d) $\alpha=1.5,h=0.6.$}
\label{fig:n_up_down}
\end{figure}

If we turn on SOC, it  shows that, for small and moderate strength of disorder, the fluctuations are significantly suppressed. The fluctuations of large $|V_i|$ and small $|V_i|$ are comparable. Although the fluctuations are suppressed, the mean order parameter is not zero but remains a finite value which is larger than the system without SOC. For large strength of disorder, the fluctuations near small $|V_i|$ are more significant than the fluctuations near large $|V_i|$. As we can see from the third column of Fig. \ref{fig:disorder}, SOC smooths the fluctuations of $\Delta_i$, and we also find that the Zeeman field has limited effect on the local order parameters $\Delta_i$.

\begin{figure}[ht]
\centering
\includegraphics[height=9cm]{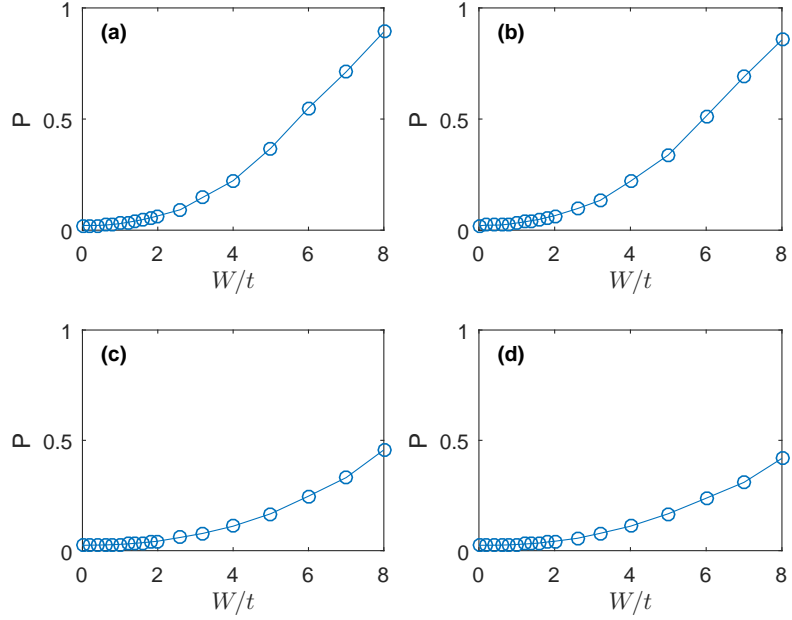}
\caption{Inverse participation ratio $P$ as a function of disorder width $W$. For localized states, $P\to1$; for extended states, $P\to 0.$ (a) $\alpha=0.75,h=0.3.$ (b) $\alpha=0.75,h=0.6.$ (c) $\alpha=1.5,h=0.3.$ (d) $\alpha=1.5,h=0.6.$}
\label{fig:ipr}
\end{figure}

All the findings show that SOC has the opposite role compared to the disorder. Disorder makes the particles more concentrated while SOC makes the particle more mobile. This is because SOC can make fermions hop between the nearest-neighbor sites with spin flipping and thus has a strong effect on the number difference. In Fig. \ref{fig:n_up_down}, we plotted the number density for spin-up ($n_{\uparrow}$) and spin-down ($n_{\downarrow}$) particles. In the absence of disorder, increasing of SOC or Zeeman field increases the difference between $n_{\uparrow}$ and $n_{\downarrow}$\cite{Zheng:2014}. This is obvious because Zeeman field causes energy level splitting between spin-up state and spin-down state. And SOC will cause the transfer between spin-up and spin-down particles. From Fig. \ref{fig:n_up_down} (a) and (c), we can see that Zeeman field is the main reason for causing the difference between the two kind of particles, increasing strength of SOC will barely change the difference\cite{Liang:2015}. If we turn on the disorder, in all cases, the difference between $n_{\uparrow}$ and $n_{\downarrow}$ will decrease. This decreasing means that the disorder neutralizes the effects caused by SOC and Zeeman field. This explains why the system with SOC is more insensitive to disorder.


As we can see from above analysis, when we increase the strength of disorder, the system will undergo a transition from a gapped superconductor to a gapped insulator. To demonstrate this effect, we calculate the inverse participation ratio (IPR) $P$ as follows
\begin{equation}
P=\frac{\sum(|u_{i\sigma}^n|^4+|v_{i\sigma}^n|^4)}{\sum(|u_{i\sigma}^n|^2+|v_{i\sigma}^n|^2)}.
\end{equation}

In Fig. \ref{fig:ipr} we ploted the IPR as a function of disorder width $W$. For small SOC, with the increase of disorder width, $P$ increases from $0$ to nearly $1$. This is a strong evidence of Anderson localization. From Fig. \ref{fig:ipr} (a) and (b), we find that Zeeman field has limited effect on the IPR, which indicates the localization is insensitive to Zeeman field. This is because Anderson localization is mostly single-particle physics. For large SOC, the maximum of IPR is significanltly smaller than $1$, this means some parts of the system are in localized states and others are in extended states. Since large SOC changes the single-particle dispersion relation and causes the particles more mobile, the system is less localized than the system without SOC. We can understand this way, in the presence of SOC, the momentum distribution for spin-up and spin-down particles are broadly extended compared to the no-SOC case\cite{ZhouX:2013,Zheng:2014}. For small SOC, the two energy bands nearly degenerate, and the two bands are both partly occupied. As increasing of SOC, the Fermi pockets in Fig. \ref{fig:band} shrinks to Fermi points, the occupation of upper Rashba band becomes less, and the momentum distribution of spin-up and spin-down particles broadens a lot. This also reveals the fact that there will exist Anderson transition even in the 2-dimensional system if SOC is presented\cite{Kohmoto:2008,Zhou:2013}. Although the system studied previously\cite{Kohmoto:2008,Zhou:2013} has no interaction, in this work we have considered the interaction, and find the similar results. These confirm the findings that SOC has opposite effect of disorder.


\section*{Conclusion}
In summary, in this paper we have studied a 2D spin-orbit coupled degenerate fermionic optical lattice system with uniformly distributed random disorder and with Zeeman field. We find that, with the increase of disorder strength, the mean order parameter $\Delta$ will vanish while the energy gap $E_g$ will persist. Meanwhile, the system undergoes a transition from a superconducting state to a gapped insulating state. We calculated the density of state to show that the energy gap will never be closed no matter how strong the disorder is. In the presence of disorder and without SOC, local superconducting order parameter $\{\Delta_i\}$ has very strong fluctuations on those sites whose local lattice disorder $|V_i|$ are small. If we turn on SOC, we find the fluctuations have been suppressed. Although the maximum of local superconducting order parameter $\{\Delta_i\}$ is reduced, the mean order parameter $\Delta$ remains a finite value which is larger than the value of system without SOC. Lastly, we calculated the inverse participation ratio, it shows that the system with SOC and disorder has a mixed state: some parts of the system are in localized states and others are in extended states. This confirms the same findings in the interaction-free system by other groups. All the findings show that SOC makes the particles more mobile and causes the opposite effect of disorder. And strong SOC will make the system more insensitive to the disorder.

\section*{Methods}
Under the mean-field approximation, the on-site interaction term can be written as
\begin{equation}
-U{\hat n}_{i\uparrow}{\hat n}_{i\downarrow}\simeq\Delta_{i}^{*}c_{i\downarrow}c_{i\uparrow}+\Delta_{i}c_{i\uparrow}^{\dagger}c_{i\downarrow}^{\dagger}+
\frac{|\Delta_i|^2}{U}+U\langle{\hat n}_{i\uparrow}\rangle{\hat n}_{i\downarrow}+U\langle{\hat n}_{i\downarrow}\rangle{\hat n}_{i\uparrow}.
\end{equation}
Here we only consider the Hartree correction term $U\langle{\hat n}_{i\uparrow}\rangle{\hat n}_{i\downarrow}+U\langle{\hat n}_{i\downarrow}\rangle{\hat n}_{i\uparrow}$ and the local order parameter is $\Delta_i=-U\langle c_{i\downarrow}c_{i\uparrow}\rangle$.

To diagonalize the Hamiltonian, we use the usual Bogoliubov-de Gennes transformation\cite{Gennes:1966}
\begin{equation}
\begin{matrix}
c_{i\uparrow}=\sum_{n=1}^{2N}(u_{i\uparrow}^{n}\Gamma_n-v_{i\uparrow}^{n*}\Gamma_{n}^{\dagger}),\\
c_{i\downarrow}=\sum_{n=1}^{2N}(u_{i\downarrow}^{n}\Gamma_n+v_{i\downarrow}^{n*}\Gamma_{n}^{\dagger}),
\end{matrix}
\end{equation}
where $\{\Gamma_n\}$ and $\{\Gamma_n^{\dagger}\}$ are the quasi-particle operators, $N$ is the number of sites.

Substitute above equations to the Hamiltonian, we can obtain
\begin{equation}
\sum_{j}\begin{pmatrix}H_{ij\uparrow\uparrow} & H_{ij\uparrow\downarrow} & \Delta_{ij} & 0 \\
H_{ij\downarrow\uparrow} & H_{ij\downarrow\downarrow} & 0 & \Delta_{ij} \\
\Delta_{ij}^{*} & 0 & -H_{ij\downarrow\downarrow}^{*} & H_{ij\downarrow\uparrow}^{*} \\
0 & \Delta_{ij}^{*} & H_{ij\uparrow\downarrow}^{*} & -H_{ij\uparrow\uparrow}^{*}
\end{pmatrix}\Psi_{j}^{n} = E_n\Psi_{i}^{n},
\end{equation}
where $H_{ij\uparrow\uparrow}=-t\delta_{\langle ij\rangle}-(\mu-V_i-h+U\langle{\hat n}_{i\downarrow}\rangle)\delta_{ij}$, $H_{ij\downarrow\downarrow}=-t\delta_{\langle ij\rangle}-(\mu-V_i+h+U\langle{\hat n}_{i\uparrow}\rangle)\delta_{ij}$. For the $x$-direction hopping, $H_{ij\uparrow\downarrow}=\pm\alpha\delta_{\langle ij \rangle}$ (`$+$' for the positive-$x$ hopping and `$-$' for the negative-$x$ hopping). For the $y$-direction hopping, $H_{ij\uparrow\downarrow}=\mp i\alpha\delta_{\langle ij \rangle}$(`$-$' for the positive-$y$ hopping and `$+$' for the negative-$y$ hopping). And we have $H_{ij\uparrow\downarrow}=H_{ij\downarrow\uparrow}^{*}$ for the $x$- and $y$-direction hopping, respectively. $\delta_{\langle ij\rangle}=1$ is used for nearest-neighbour sites and $\delta_{\langle ij\rangle}=0$ for others. The quasi-particle wave function is $\Psi_{j}^{n}=(u_{j\uparrow}^n,u_{j\downarrow}^n,v_{j\downarrow}^n,v_{j\uparrow}^n)^T$.

We solve the Bogoliubov-de Gennes equation self-consistently at temperature $T=0$. The self-consistence equations consist of the number equation
\begin{equation}
\langle {\hat n}_{i\sigma}\rangle=\sum_{n=1}^{2N}[|u_{i\sigma}^{n}|^2f(E_n)+|v_{i\sigma}^{n}|^2f(-E_n)],
\end{equation}
and the gap equation
\begin{equation}
\Delta_{ij}=U\delta_{ij}\sum_{n=1}^{2N}\{u_{i\uparrow}^{n}v_{i\downarrow}^{n*}f(-E_n)-u_{i\downarrow}^{n}v_{i\uparrow}^{n*}f(E_n)\}.
\end{equation}

Here
$f(E_n)$ is the Fermi-Dirac distribution. Since we only consider the case of $T=0$, we have
\[f(E_n) =
  \begin{cases}
    0  & \quad \text{if } E_n  > 0,\\
    1  & \quad \text{if } E_n < 0.\\
  \end{cases}
\]


\begin{thebibliography}{10}
\expandafter\ifx\csname url\endcsname\relax
  \def\url#1{\texttt{#1}}\fi
\expandafter\ifx\csname urlprefix\endcsname\relax\def\urlprefix{URL }\fi
\providecommand{\bibinfo}[2]{#2}
\providecommand{\eprint}[2][]{\url{#2}}

\bibitem{Mott:1949}
\bibinfo{author}{Mott, N.~F.}
\newblock \bibinfo{title}{The basis of the electron theory of metals, with
  special reference to the transition metals}.
\newblock \emph{\bibinfo{journal}{Proceedings of the Physical Society. Section
  A}} \textbf{\bibinfo{volume}{62}}, \bibinfo{pages}{416}
  (\bibinfo{year}{1949}).

\bibitem{Anderson:1958}
\bibinfo{author}{Anderson, P.~W.}
\newblock \bibinfo{title}{Absence of diffusion in certain random lattices}.
\newblock \emph{\bibinfo{journal}{Phys. Rev.}} \textbf{\bibinfo{volume}{109}},
  \bibinfo{pages}{1492--1505} (\bibinfo{year}{1958}).

\bibitem{Abrahams:1979}
\bibinfo{author}{Abrahams, E.}, \bibinfo{author}{Anderson, P.~W.},
  \bibinfo{author}{Licciardello, D.~C.} \& \bibinfo{author}{Ramakrishnan,
  T.~V.}
\newblock \bibinfo{title}{Scaling theory of localization: Absence of quantum
  diffusion in two dimensions}.
\newblock \emph{\bibinfo{journal}{Phys. Rev. Lett.}}
  \textbf{\bibinfo{volume}{42}}, \bibinfo{pages}{673--676}
  (\bibinfo{year}{1979}).

\bibitem{Shapiro:2012}
\bibinfo{author}{Shapiro, B.}
\newblock \bibinfo{title}{Cold atoms in the presence of disorder}.
\newblock \emph{\bibinfo{journal}{J. of Phys. A: Mathematical and Theoretical}}
  \textbf{\bibinfo{volume}{45}}, \bibinfo{pages}{143001}
  (\bibinfo{year}{2012}).

\bibitem{Semeghini:2015}
\bibinfo{author}{Semeghini, G.} \emph{et~al.}
\newblock \bibinfo{title}{Measurement of the mobility edge for 3d anderson
  localization}.
\newblock \emph{\bibinfo{journal}{Nat. Phys.}} \textbf{\bibinfo{volume}{11}},
  \bibinfo{pages}{554--559} (\bibinfo{year}{2015}).

\bibitem{Sanchez:2007}
\bibinfo{author}{Sanchez-Palencia, L.} \emph{et~al.}
\newblock \bibinfo{title}{Anderson localization of expanding bose-einstein
  condensates in random potentials}.
\newblock \emph{\bibinfo{journal}{Phys. Rev. Lett.}}
  \textbf{\bibinfo{volume}{98}}, \bibinfo{pages}{210401}
  (\bibinfo{year}{2007}).

\bibitem{Billy:2008}
\bibinfo{author}{Billy, J.} \emph{et~al.}
\newblock \bibinfo{title}{Direct observation of anderson localization of matter
  waves in a controlled disorder}.
\newblock \emph{\bibinfo{journal}{Nature}} \textbf{\bibinfo{volume}{453}},
  \bibinfo{pages}{891--894} (\bibinfo{year}{2008}).


\bibitem{Cao:2015}
\bibinfo{author}{Cao, Y.,} \bibinfo{author}{Gao, X.,} \bibinfo{author}{Liu, X.,} \bibinfo{author}{Pu, H.}
\newblock \bibinfo{title}{Anderson localization of Cooper pairs and Majorana fermions in an ultracold atomic Fermi gas with synthetic spin-orbit coupling}.
\newblock \emph{\bibinfo{journal}{arXiv: 1512.03447v1}}
  (\bibinfo{year}{2015}).

\bibitem{Modugno:2010}
\bibinfo{author}{Modugno, G.}
\newblock \bibinfo{title}{Anderson localization in bose–einstein
  condensates}.
\newblock \emph{\bibinfo{journal}{Rep. Prog. Phys.}}
  \textbf{\bibinfo{volume}{73}}, \bibinfo{pages}{102401}
  (\bibinfo{year}{2010}).

\bibitem{Cai:2013}
\bibinfo{author}{Cai, X.}, \bibinfo{author}{Lang, L.-J.},
  \bibinfo{author}{Chen, S.} \& \bibinfo{author}{Wang, Y.}
\newblock \bibinfo{title}{Topological superconductor to anderson localization
  transition in one-dimensional incommensurate lattices}.
\newblock \emph{\bibinfo{journal}{Phys. Rev. Lett.}}
  \textbf{\bibinfo{volume}{110}}, \bibinfo{pages}{176403}
  (\bibinfo{year}{2013}).

\bibitem{Cui:2008}
\bibinfo{author}{Cui, Q.} \& \bibinfo{author}{Yang, K.}
\newblock \bibinfo{title}{Fulde-ferrell-larkin-ovchinnikov state in disordered
  $s$-wave superconductors}.
\newblock \emph{\bibinfo{journal}{Phys. Rev. B}} \textbf{\bibinfo{volume}{78}},
  \bibinfo{pages}{054501} (\bibinfo{year}{2008}).

\bibitem{Dubi:2007}
\bibinfo{author}{Dubi, Y.}, \bibinfo{author}{Meir, Y.} \&
  \bibinfo{author}{Avishai, Y.}
\newblock \bibinfo{title}{Nature of the superconductor-insulator transition in
  disordered superconductors}.
\newblock \emph{\bibinfo{journal}{Nature}} \textbf{\bibinfo{volume}{449}},
  \bibinfo{pages}{876--880} (\bibinfo{year}{2007}).

\bibitem{Dubi:2008}
\bibinfo{author}{Dubi, Y.}, \bibinfo{author}{Meir, Y.} \&
  \bibinfo{author}{Avishai, Y.}
\newblock \bibinfo{title}{Island formation in disordered superconducting thin
  films at finite magnetic fields}.
\newblock \emph{\bibinfo{journal}{Phys. Rev. B}} \textbf{\bibinfo{volume}{78}},
  \bibinfo{pages}{024502} (\bibinfo{year}{2008}).

\bibitem{Ghosal:1998}
\bibinfo{author}{Ghosal, A.}, \bibinfo{author}{Randeria, M.} \&
  \bibinfo{author}{Trivedi, N.}
\newblock \bibinfo{title}{Role of spatial amplitude fluctuations in highly
  disordered $\mathit{s}$-wave superconductors}.
\newblock \emph{\bibinfo{journal}{Phys. Rev. Lett.}}
  \textbf{\bibinfo{volume}{81}}, \bibinfo{pages}{3940--3943}
  (\bibinfo{year}{1998}).

\bibitem{Ghosal:2001}
\bibinfo{author}{Ghosal, A.}, \bibinfo{author}{Randeria, M.} \&
  \bibinfo{author}{Trivedi, N.}
\newblock \bibinfo{title}{Inhomogeneous pairing in highly disordered \textit{s}
  -wave superconductors}.
\newblock \emph{\bibinfo{journal}{Phys. Rev. B}} \textbf{\bibinfo{volume}{65}},
  \bibinfo{pages}{014501} (\bibinfo{year}{2001}).

\bibitem{He:2013}
\bibinfo{author}{He, L.} \& \bibinfo{author}{Song, Y.}
\newblock \bibinfo{title}{Self-consistent calculations of the effects of
  disorder in d-wave and s-wave superconductors}.
\newblock \emph{\bibinfo{journal}{J. of the Korean Phys. Soc.}}
  \textbf{\bibinfo{volume}{62}}, \bibinfo{pages}{2223--2227}
  (\bibinfo{year}{2013}).

\bibitem{Sacepe:2011}
\bibinfo{author}{Sacepe, B.} \emph{et~al.}
\newblock \bibinfo{title}{Localization of preformed cooper pairs in disordered
  superconductors}.
\newblock \emph{\bibinfo{journal}{Nat. Phys.}} \textbf{\bibinfo{volume}{7}},
  \bibinfo{pages}{239--244} (\bibinfo{year}{2011}).

\bibitem{Roati:2008}
\bibinfo{author}{Roati, G.} \emph{et~al.}
\newblock \bibinfo{title}{Anderson localization of a non-interacting
  bose-einstein condensate}.
\newblock \emph{\bibinfo{journal}{Nature}} \textbf{\bibinfo{volume}{453}},
  \bibinfo{pages}{895--898} (\bibinfo{year}{2008}).

\bibitem{Potirniche:2014}
\bibinfo{author}{Potirniche, I.-D.}, \bibinfo{author}{Maciejko, J.},
  \bibinfo{author}{Nandkishore, R.} \& \bibinfo{author}{Sondhi, S.~L.}
\newblock \bibinfo{title}{Superconductivity of disordered dirac fermions in
  graphene}.
\newblock \emph{\bibinfo{journal}{Phys. Rev. B}} \textbf{\bibinfo{volume}{90}},
  \bibinfo{pages}{094516} (\bibinfo{year}{2014}).

\bibitem{Nanguneri:2012}
\bibinfo{author}{Nanguneri, R.} \emph{et~al.}
\newblock \bibinfo{title}{Interplay of superconductivity and spin-dependent
  disorder}.
\newblock \emph{\bibinfo{journal}{Phys. Rev. B}} \textbf{\bibinfo{volume}{85}},
  \bibinfo{pages}{134506} (\bibinfo{year}{2012}).

\bibitem{Jiang:2013}
\bibinfo{author}{Jiang, M.} \emph{et~al.}
\newblock \bibinfo{title}{Gapless inhomogeneous superfluid phase with
  spin-dependent disorder}.
\newblock \emph{\bibinfo{journal}{New J. of Phys.}}
  \textbf{\bibinfo{volume}{15}}, \bibinfo{pages}{023023}
  (\bibinfo{year}{2013}).

\bibitem{Sanchez:2008}
\bibinfo{author}{Sanchez-Palencia, L.} \emph{et~al.}
\newblock \bibinfo{title}{Disorder-induced trapping versus anderson
  localization in bose–einstein condensates expanding in disordered
  potentials}.
\newblock \emph{\bibinfo{journal}{New J. of Phys.}}
  \textbf{\bibinfo{volume}{10}}, \bibinfo{pages}{045019}
  (\bibinfo{year}{2008}).

\bibitem{Zhou:2013}
\bibinfo{author}{Zhou, L.}, \bibinfo{author}{Pu, H.} \& \bibinfo{author}{Zhang,
  W.}
\newblock \bibinfo{title}{Anderson localization of cold atomic gases with
  effective spin-orbit interaction in a quasiperiodic optical lattice}.
\newblock \emph{\bibinfo{journal}{Phys. Rev. A}} \textbf{\bibinfo{volume}{87}},
  \bibinfo{pages}{023625} (\bibinfo{year}{2013}).

\bibitem{Trivedi:2012}
\bibinfo{author}{Trivedi, N.,} \bibinfo{author}{Loh, Y.,} \bibinfo{author}{Bouadim, K.,} \bibinfo{author}{Randeria, M.}
\bibinfo{title}{Aspects of localization across the 2D superconductor-insulator transition}.
\emph{\bibinfo{journal}{Int. J. of Modern Physics: Conference Series.}} \textbf{\bibinfo{volume}{11}}, \bibinfo{pages}{22-37}
(\bibinfo{year}{2012}).


\bibitem{Chen:2012}
\bibinfo{author}{Chen, G.,} \bibinfo{author}{Gong, M.,}\bibinfo{author}{Zhang, C.}
\bibinfo{title}{BCS-BEC crossover in spin-orbit-coupled two-dimensional Fermi gases}.
\emph{\bibinfo{journal}{Phys. Rev. A}} \textbf{\bibinfo{volume}{85}}, \bibinfo{pages}{013601}
(\bibinfo{year}{2012}).

\bibitem{Gennes:1966}
\bibinfo{author}{de~Gennes, P.~G.}
\newblock \emph{\bibinfo{journal}{Superconductivity in Metals and Alloys,}}
  \bibinfo{pages}{Westview Press,} (\bibinfo{year}{March 31, 1999}).

\bibitem{Bouadim:2011}
\bibinfo{author}{Bouadim, K.}, \bibinfo{author}{Loh, Y.~L.},
  \bibinfo{author}{Randeria, M.} \& \bibinfo{author}{Trivedi, N.}
\newblock \bibinfo{title}{Single- and two-particle energy gaps across the
  disorder-driven superconductor-insulator transition}.
\newblock \emph{\bibinfo{journal}{Nat. Phys.}} \textbf{\bibinfo{volume}{7}},
  \bibinfo{pages}{884--889} (\bibinfo{year}{2011}).

\bibitem{Kohmoto:2008}
\bibinfo{author}{Kohmoto, M.} \& \bibinfo{author}{Tobe, D.}
\newblock \bibinfo{title}{Localization problem in a quasiperiodic system with
  spin-orbit interaction}.
\newblock \emph{\bibinfo{journal}{Phys. Rev. B}} \textbf{\bibinfo{volume}{77}},
  \bibinfo{pages}{134204} (\bibinfo{year}{2008}).

\bibitem{Zheng:2014}
\bibinfo{author}{Zheng, Z.} \emph{et~al.}
\newblock \bibinfo{title}{FFLO superfluids in 2D spin-orbit Coupled Fermi gases}.
\newblock \emph{\bibinfo{journal}{Sci. Rep.}} \textbf{\bibinfo{volume}{4}},
  \bibinfo{pages}{6535} (\bibinfo{year}{2014}).

\bibitem{Liang:2015}
\bibinfo{author}{Liang, J.} \emph{et~al.}
\newblock \bibinfo{title}{Unconventional pairing of spin-orbit coupled attractive degenerate Fermi gas in a one-dimensional optical lattice}.
\newblock \emph{\bibinfo{journal}{Sci. Rep.}} \textbf{\bibinfo{volume}{5}},
  \bibinfo{pages}{14863} (\bibinfo{year}{2015}).

\bibitem{Xu:2014}
\bibinfo{author}{Xu, Y.}, \bibinfo{author}{Qu, C.}, \bibinfo{author}{Gong, M.}, \& \bibinfo{author}{Zhang, C.}
\newblock \bibinfo{title}{Competing superfluid orders in spin-orbit-coupled fermionic cold-atom optical lattices}.
\newblock \emph{\bibinfo{journal}{Phys. Rev. A}} \textbf{\bibinfo{volume}{89}}, \bibinfo{pages}{013607}
  (\bibinfo{year}{2014}).

\bibitem{Sun:2013}
\bibinfo{author}{Sun, Q.,} \bibinfo{author}{Zhu, G.,} \bibinfo{author}{Liu, W.,} \& \bibinfo{author}{Ji, A.}
\bibinfo{title}{Spin-orbit coupling effects on the superfluidity of a Fermi gas in an optical lattice}.
\emph{\bibinfo{journal}{Phys. Rev. A}} \textbf{\bibinfo{volume}{88}}, \bibinfo{pages}{063637}
(\bibinfo{year}{2013}).

\bibitem{ZhouX:2013}
\bibinfo{author}{Zhou, X.} \emph{et~al.}
\bibinfo{title}{Spin-orbit coupled replusive Fermi atoms in a one-dimensional optical lattice}.
\emph{\bibinfo{journal}{New J. Phys.}} \textbf{\bibinfo{volume}{17}}, \bibinfo{pages}{093044}
(\bibinfo{year}{2015}).
\end{thebibliography}


\section*{Acknowledgements}
This work is supported by National Natural Science Foundation of China (No. 61275122 and No. 61590932) and Strategic Priority Research Program (B) of CAS (No. XDB01030200).

\section*{Author contributions statement}
S.L., Y.Z. and X.Z. proposed the idea. S.L. carried out the numerical simulations and wrote the manuscript. S.L., Y.Z., and X.Z did the theoretical analysis. Y.Z. and G.G. supervised the whole research project.

\section*{Additional Information}
\textbf{Competing financial interests:} The authors declare no competing financial interests.
\end{document}